\documentclass[prd,onecolumn,showpacs,superscriptaddress,nofootinbib,floatfix,12pt,longbibliography]{revtex4-1} 
\usepackage[utf8]{inputenc}
\usepackage{color}
\usepackage{amsfonts,amsmath,amssymb}
\usepackage{graphicx}
\usepackage{multirow}
\usepackage{comment}
\usepackage{tablefootnote}
\usepackage{siunitx}
\usepackage{pifont}
\usepackage{slashed}
\usepackage{gensymb}
\usepackage{bm}
\usepackage{hyperref}
\usepackage{cleveref}
\hypersetup{colorlinks = true,
           allcolors = blue}
\usepackage{epstopdf}
\usepackage{makecell}
\usepackage[dvipsnames]{xcolor}
\usepackage{tikz}
\usepackage{dsfont}     
\usetikzlibrary{decorations.pathmorphing}
\usepackage{subfigure}
\usepackage{circuitikz}
\usepackage{tkz-graph}
\usetikzlibrary{arrows}

\def\x4c/{$X_{cc\bar c \bar c }$}
\def\xyy/{$\psi(4360)$}
\def\ppsi/{$\psi(4415)$}
\def\doned/{$D_1\bar{D}$}
\def\ddstar/{$D^* \bar{D}^*$}
\def\dordstar/{$D^{(*)} \bar{D}^{(*)}$}
\def\dtwodstar/{$D_2^*\bar{D}^*$} 
\def\dd/{$D \bar{D}$}
\def\*{^{(*)}}
\def\>{\big>}
\def\<{\big<}
\def\|{\big\vert}

\def\s{\boldsymbol{\sigma}}
\def\l{\boldsymbol{\lambda}}
\def\three{\mathbf 3}
\def\one{\mathbf 1}
\def\eight{\mathbf 8}
\def\six{\mathbf 6}

\def\c{\bar c}
\def\b{\bar b}
\def\q{\bar q}
\def\Q{\bar Q}
\def\f{\varphi}
\def\jp{J/\psi}
\def\M{\overline M}
\def\D{\bar D}
\def\*{^{(*)}}

\newcommand{\midarrow}{\tikz \draw[very thick, ->] (0,0) -- +(.2,0);}

\def\be{\begin{equation}}
\def\ee{\end{equation}}
\def\ba{\begin{eqnarray}}
\def\ea{\end{eqnarray}}

\renewcommand{\arraystretch}{1.25}

\begin{document}

\author{Muhammad Naeem Anwar}\email{m.n.anwar@swansea.ac.uk}
\author{Timothy J. Burns}\email{t.burns@swansea.ac.uk}
\affiliation{Department of Physics, Swansea University, Singleton Park, Swansea, SA2 8PP, UK.}


\title{Structure of $cc\c\c$ tetraquarks and interpretation of LHC states}

\begin{abstract}

Motivated by recent experimental evidence for apparent $cc\c\c$ states at LHCb, CMS and ATLAS, we consider how the mass spectrum and decays of such states can be used to discriminate among their possible theoretical interpretations, with a particular focus on identifying whether quarks or diquarks are the most relevant degrees of freedom. Our preferred scenario is that $X(6600)$ and its apparent partner state $X(6400)$ are the tensor $(2^{++})$ and scalar $(0^{++})$ states of an S-wave multiplet of $cc\c\c$ states. Using tetraquark mass relations which are independent of (or only weakly dependent on) model parameters, we give predictions for the masses of additional partner states with axial and scalar quantum numbers. Additionally, we give predictions for relations among decay branching fractions to $\jp\jp$, $\jp\eta_c$, $\eta_c\eta_c$ and $D\*\D\*$ channels. The scenario we consider is consistent with existing experimental data on $\jp\jp$, and our predictions for partner states and their decays can be confronted with future experimental data, to discriminate between quark and diquark models. 

\end{abstract}

\date{\today}
\maketitle

\vspace{2cm}

\section{Introduction}

Among exotic multiquark states, those with exclusively heavy quarks -- such as $cc \bar c \bar c$ and the bottom analogue $bb \bar b \bar b$ -- are particularly interesting since, owing to the absence of light degrees of freedom, they are useful to investigate the interplay between the perturbative and nonperturbative regimes of Quantum Chromodynamics (QCD) and provide a useful platform to investigate the low-energy dynamics of QCD~\cite{Anwar:2017toa,Chao:2020dml}. There is a considerable body of literature in which such states have been predicted, in a range of theoretical models, including the constituent quark model with one-gluon-exchange (OGE) interaction~\cite{Weinstein:1983gd,Zhang:2022qtp,Liu:2019zuc,Wang:2022yes,An:2022qpt,Wang:2019rdo,Anwar:2018sol,Jin:2020jfc,Lu:2020cns,Gordillo:2020sgc,Lloyd:2003yc,liu:2020eha,Yu:2022lak},
the chromomagnetic quark model~\cite{Buccella:2006fn,Deng:2020iqw,Wu:2016vtq,Weng:2020jao,Liu:2019zoy},
and the diquark model~\cite{Karliner:2016zzc,Debastiani:2017msn,Dong:2022sef,Bedolla:2019zwg,Faustov:2020qfm,Bedolla:2019zwg,Giron:2020wpx,Lundhammar:2020xvw,Berezhnoy:2011xn,Esposito:2018cwh,Karliner:2020dta,Karliner:2020dta,Sonnenschein:2020nwn,Mutuk:2021hmi}.
All of these studies focus on the mass spectrum, except for a few~\cite{Chao:1980dv,*Chao:1979mm,Chen:2020xwe,Becchi:2020uvq,Chen:2022sbf,Agaev:2023wua,Wang:2023kir} which also address decays.

The experimental era of all-heavy tetraquark spectroscopy started at LHCb in 2020, with the first observation of an apparent $cc \bar c \bar c$ state, dubbed $X(6900)$, in the $J/\psi J/\psi$ final state~\cite{LHCb:2020bwg}. Model scenarios were then considered in, for example, Refs.~\cite{Chao:2020dml,Maiani:2020pur,Richard:2020hdw,Karliner:2020dta}. The $X(6900)$ state was subsequently confirmed at CMS which, in addition, identified two further states in $J/\psi J/\psi$ decays, reported as $X(6600)$ and $X(7300)$~\cite{CMS:2023owd}. At ATLAS, the state $X(6900)$ was confirmed in $J/\psi J/\psi$ and $J/\psi \psi(2S)$, and a significant excess around the $X(7300)$ mass region was found~\cite{ATLAS:2023bft}; their extracted parameters for $X(6600)$ agree with the CMS results.
Interestingly, there is a hint in the CMS data~\cite{CMS:2023owd} that there could be an additional state around $6400$~MeV, and we refer to this as $X(6400)$. Moreover, the ATLAS data~\cite{ATLAS:2023bft} also show a similar peak structure around this  region, with mass $6410 \pm 80$ MeV. Different scenarios for interpretation of these states as $cc\c\c$ tetraquarks were considered in Refs.~\cite{Lu:2020cns,liu:2020eha,Deng:2020iqw,Zhang:2022qtp,Wang:2022yes,Yu:2022lak,Maiani:2022psl,Dong:2022sef,An:2022qpt,Ortega:2023pmr,Wang:2023jqs}.

A brief summary of extracted parameters of $cc \bar c \bar c$ states by different LHC experiments is given in Table\,\ref{tab:expdata}. Despite some differences in the parameters, there is a clear consensus for the existence of several peaks/dip(s) in the mass region $(6.2 \sim 7.5)$ GeV in both $J/\psi J/\psi$ and $J/\psi \psi(2S)$ final states. In this paper we compare this emerging body of experimental data on $cc\c\c$ states to the predictions of diverse theoretical approaches, aiming to identify and discriminate among various plausible model scenarios.

As well as the experiments at the LHC, the future Super $\tau$-Charm Facility STCF~\cite{Achasov:2023gey}, which is currently under development, will be ideal for the study of $cc\c\c$ states. The center-of-mass energy of this electron-positron collider can reach 7~GeV, which is sufficient for the production of two $c\c$ pairs, and covers the relevant mass range of the $cc\c\c$ states discovered so far, and their presumed partners. In addition to decays into charmonia pairs (such as $J/\psi J/\psi$), one also expects $cc\c\c$ states to decay into pairs of charm and anti-charm mesons (such as $D\*\D\*$) via the annihilation of a $c\c$ pair into a gluon. Identifying such decays at the LHC will be difficult, due to the high background. Hence, the STCF will be an ideal place to establish the existence of all-charm tetraquarks by searching for them in different final states. 

\begin{table}[t!]
\begin{center}
\begin{tabular}{p{2cm} p{2.2cm} p{2.5cm} p{2.5cm} p{2.5cm}}
\hline\hline
State & Parameters & LHCb~\cite{LHCb:2020bwg} & CMS~\cite{CMS:2023owd} & ATLAS~\cite{ATLAS:2023bft} \\ [1ex]
    \hline
    \multirow{2}{*}{$X(6900)$}& M (MeV) & $6905 \pm 11 \pm 7$  & $6927 \pm 9 \pm 4$ & $6860 \pm 30^{+10}_{-20}$\\
    & $\Gamma$ (MeV) & $80 \pm 19 \pm 33$ &  $122^{+24}_{-21} \pm 18$ & $110 \pm 50^{+20}_{-10}$ \\
    \hline
    \multirow{2}{*}{$X(6600)$}& M (MeV) &   & $6552 \pm 10 \pm 12$ & $6630 \pm 50^{+80}_{-10}$\\
    & $\Gamma$ (MeV) &  & $124^{+32}_{-26} \pm 33$ & $350 \pm 110^{+110}_{-40}$ \\
     \hline
    \multirow{2}{*}{$X(6400)$}& M (MeV) &   & $(6402 \pm 15)^{\dag}$ & $6410 \pm 80^{+80}_{-30}$\\
    & $\Gamma$ (MeV) &  &  & $590 \pm 350^{+120}_{-200}$ \\
    \hline
\end{tabular}
\caption{Masses and decay widths of $cc \bar c \bar c$ states extracted by different LHC experiments in $J/\psi J/\psi$ mass spectrum.
$\dag$ This entry is based on our finding that there should be another (small) peak in the CMS data~\cite{CMS:2023owd}, which we spot around 6400 MeV. More details will be discussed in the text.}
\end{center}
\label{tab:expdata}
\end{table}

We recently derived a number of general results for the spectrum of S-wave tetraquarks with either two flavours ($QQ\q\q$) or one ($QQ\Q\Q$)~\cite{Anwar:2023svj}, the latter case of course being of interest to the present work on $cc\c\c$ states. We found results which apply to both quark and diquark models (which have characteristically different colour wavefunctions) and also to different variants of each of model, with either effective (di)quark masses, or dynamical masses obtained from the Schr\"{o}dinger equation. In particular we derived mass formulae which we will use, in this paper, to inform our preferred assignment of quantum numbers to the experimental candidates. In  Ref.~\cite{Anwar:2023svj} we also identified new linear relations among tetraquark masses which we will apply, in the current work, to predict the masses of  partner states which have yet to be discovered; these predictions have either no dependence, or only a very weak dependence, on model parameters. We also derived results on the colour mixing which we will use, in this paper, to predict the relative decay rates of $cc\c\c$ states to different final states.

In Section\,\ref{sec:general} we discuss some general features of the spectroscopy of $cc\c\c$ states, and suggest a scenario in which $X(6600)$, and an apparent experimental signal which we refer to as $X(6400)$, are the $2^{++}$ and $0^{++}$ states in the ground state S-wave multiplet of $cc\c\c$ states. Drawing on the results of our recent paper~\cite{Anwar:2023svj}, in Section\,\ref{sec:spectrum} we present general formulae for the mass spectra of $cc\c\c$ states in quark and diquark models. In Section\,\ref{sec:interpretation} we compare these results to the experimental candidates, and predict the masses of additional partner states which have yet to be discovered in experiment, considering also the extent of model dependence in these predictions.
In Section\,\ref{sec:decays} we give predictions for the relative partial widths of $cc\c\c$ states to different charmonia (such as $\jp\jp$ and $\eta_c\eta_c$), and different combinations of open charm mesons ($D\*\D\*$), and show how experimental observation of these decays can discriminate among models. Finally, conclusions and outlook are given in Section\,\ref{sec:concl}.

\section{General features}
\label{sec:general}

The quantum numbers of the ground state multiplet of $cc\c\c$ states are fixed by the Pauli principle, which constrains the colour and spin configurations of the $cc$ and $\c\c$ pairs. In a relative $S$-wave, a $cc$ pair can have (colour, spin) quantum numbers ($\bar\three$,1) or ($\six$,0), while a $\c\c$ pair can be  ($\three$,1) or ($\bar\six$,0). Combining the spins in S-wave to angular momentum $J$, and the colours to form a colour singlet, the allowed combinations (and their $J^{PC}$ quantum numbers) are
\begin{align}
\|\f_2\>&=\|\{(cc)_{\bar\three}^1(\c\c)_\three^1\}^2\> \quad (2^{++}),\label{eq:basis2}\\
\|\f_1\>&=\|\{(cc)_{\bar\three}^1(\c\c)_\three^1\}^1\> \quad (1^{+-}),\label{eq:basis1}\\
\|\f_0\>&=\|\{(cc)_{\bar\three}^1(\c\c)_\three^1\}^0\> \quad (0^{++}),\label{eq:basis0}\\
\|\f_0'\>&=|\{(cc)_{\six}^0(\c\c)_{\bar\six}^0\}^0\> \quad (0^{++}),\label{eq:basis0p}
\end{align}
where on the right-hand side, the subscripts are colour, and superscripts are spin.

A basic assumption of diquark models is that states are built out of the (hidden) colour triplet configurations only, so the spectrum has three states $\f_2$, $\f_1$ and $\f_0$, with distinct quantum numbers. Quark models, by contrast, include both the colour triplet and colour sextet combinations, so there are two scalar states, which we will refer to as $0^{++}$ and ${0^{++}}'$, which are admixtures of $\f_0$ and $\f_0'$. Obviously, experimental determination of the number of scalar states in the mass spectrum can immediately discriminate between quark models (two states) and diquark models (one). 

The allowed decays of $cc\c\c$ states to  combinations of $\jp$ and $\eta_c$ are constrained by charge conjugation symmetry. The channels accessible in S-wave are
\begin{align}
    2^{++}&\to \jp\jp \, , \label{eq:rearr1}\\
    1^{+-}&\to \jp\eta_c \, ,\label{eq:rearr2}\\
    0^{++(\prime)}&\to \jp\jp,~\eta_c\eta_c \, .\label{eq:rearr3}
\end{align}
The $2^{++}$ state can also decay to $\eta_c\eta_c$ in D-wave, but due to the centrifugal factor in the decay amplitude we assume this is comparatively insignificant.

Because the experimental states are seen in $\jp\jp$, their possible quantum numbers are $0^{++}$ or $2^{++}$. Naively we may hope that by  counting the number of peaks in the $\jp\jp$ spectrum, we could distinguish between diquark models (two peaks) and quark models (three). Indeed, with reference to Table\,\ref{tab:expdata}, it is tempting to assign all three of the states seen at ATLAS to the S-wave multiplet, and to argue in favour of the quark model on this basis; unfortunately the mass splitting in this scenario is implausibly large (see below). In any case, as we show later, not all the peaks are expected to be equally prominent in $\jp\jp$.

In Table\,\ref{tab:modelmasses} we compile some model predictions for the masses of the states in the S-wave ground state multiplet of $cc\c\c$ states. Even among models which are basically similar,  there is a very large variation in the predicted masses (and mass splittings). In some cases the predictions compare rather favourably to the experimental candidates, while in other cases the predictions are very different (generally lower). Clearly there is no prospect of assigning  quantum numbers to the states, nor of arguing in favour of one particular model, on the basis of these mass predictions alone.

\begin{table}[h!]
\def\arraystretch{1}
\begin{center}
\begin{tabular}{p{4cm} p{1cm} p{1.5cm} p{1.5cm} p{1.5cm} p{2cm} p{1.8cm} p{1.5cm}}
\hline\hline
Models && $0^{++}$&$1^{+-}$&$2^{++}$&$0^{++\prime}$ & $M_2-M_0$&$M_0'-M_0$\\ \hline
Diquark potential&\cite{Berezhnoy:2011xn}&5966 & 6051 & 6223 & & 257\\
\quad model&\cite{Faustov:2020qfm}&6190 & 6271 & 6367 && 177\\
&\cite{Lundhammar:2020xvw} &5960 & 6009 & 6100 && 140\\
\hspace{2cm}&\cite{Bedolla:2019zwg}&5883 & 6120 & 6246 & & 363\\
\hspace{2cm}&\cite{Debastiani:2017msn}&5969.4 & 6020.9 & 6115.4 & & 146\\
\hspace{2cm}&\cite{Dong:2022sef}$^{\dag}$ &6053 & 6181 & 6331 & & 278\\
\hline
Chromomagnetic&\cite{Wu:2016vtq} & 6797 & 6899 & 6956 & 7016 & 159 & 219\\
 \quad quark model&\cite{Weng:2020jao} & 6044.9 & 6230.6 & 6287.3 & 6271.3 & 242.4 & 226.4\\
\hspace{2cm}&\cite{Deng:2020iqw} & 6035 & 6139 & 6194 & 6254$^{\ddag}$ & 159 & 219 \\
\hline
Quark potential&\cite{Zhang:2022qtp} &6411 & 6453 & 6475 & 6500 & 64 & 89\\
\quad  model&\cite{Liu:2019zuc} &6455 & 6500 & 6524 & 6550 & 69 & 95\\
    \hspace{2cm}&\cite{Wang:2019rdo} &6377& 6425 & 6432 & 6425 & 55 & 48\\
\hspace{2cm}&\cite{Lu:2020cns}&6435& 6441 & 6515 & 6543 & 80 & 108\\
\hspace{2cm}&\cite{Lloyd:2003yc}&6477& 6528 & 6573 & 6695 & 96 & 218\\
\hspace{2cm}&\cite{Gordillo:2020sgc}& 6351& 6441 & 6471 & & 120 & \\

\hline
\end{tabular}
 \caption{Masses (in MeV) of the S-wave ground state $cc\c\c$ multiplet in various models, and (in the last two columns) the corresponding mass splittings. We have only included models of the type discussed in our previous paper~\cite{Anwar:2023svj}; examples of other types of models are discussed in the text. $^{\dag}$Ref.~\cite{Dong:2022sef} gives predictions with various different potential models; here we quote their results for the Godfrey-Isgur (GI) model. $^{\ddag}$Ref.~\cite{Deng:2020iqw} does not quote a prediction for the $0^{++\prime}$ state; we thank the authors for providing this in correspondence. }
\end{center}
\label{tab:modelmasses}
\end{table}

A feature common to all models, though, is that the splittings are considerably smaller than would be needed to accommodate all three candidates $X(6400)$, $X(6600)$ and $X(6900)$ in a single S-wave multiplet (as mentioned earlier). We therefore narrow our remit, and concentrate on the lower  states  $X(6400)$ and $X(6600)$, noting (Table\,\ref{tab:modelmasses}) that their masses are generally much closer to model predictions than the heavier state $X(6900)$.

As further justification for concentrating on the lower states, we note that an as alternative to the model predictions in Table\,\ref{tab:modelmasses}, we may estimate very roughly the expected masses of $cc\c\c$ states on the basis of a comparison to the recently-discovered $ccu$ baryon  $\Xi_{cc}^{++}$. In the baryon, the $cc$ pair has the same ($\bar\three$,1) quantum numbers of (colour, spin) as the $cc$ pair in the diquark model for $cc\c\c$. From the $\Xi_{cc}^{++}$ mass $3621.40 \pm 0.78 $ MeV~\cite{LHCb:2017iph}, we would guess an effective mass of around 3290~MeV for the $cc$ spin-1 diquark, where here we have attributed 330~MeV to the mass of the light quark, as is typical (see, for example, Refs.~\cite{Lichtenberg:1996fi,Lu:2016mbb}). A somewhat more intricate fit to the $cc$ diquark mass gives $3204.1$ MeV~\cite{Karliner:2016zzc}. The expected mass scale of  $cc\c\c$ ground states can be estimated, very roughly, by doubling the $cc$ diquark mass, and on this basis we notice that $X(6400)$ and $X(6600)$ masses are in the right ball park~\cite{Anwar2023} (though of course we are ignoring potentially significant contributions due to binding and spin-dependent splittings).

As is apparent in Table\,\ref{tab:modelmasses}, the masses $M_0$, $M_1$ and $M_2$ of the $0^{++}$, $1^{+-}$ and $2^{++}$ states in diquark models are ordered
\begin{align}
    M_0<M_1<M_2\, ,\label{eq:ordering1}
\end{align}
and this can be understood in general terms~\cite{Anwar:2023svj}. Noting that only 
the scalar and tensor states can decay to $\jp\jp$, then in diquark models the
$X(6400)$ and $X(6600)$ states would be assigned $0^{++}$ and $2^{++}$ quantum numbers, respectively.

The quantum number assignments are not so clear in quark models, in which there are three possible states ($0^{++}$,  $2^{++}$, ${0^{++}}'$) which decay to $\jp\jp$, and only two experimental candidates. Moreover, the relative mass $M_0'$ of the heavier scalar ${0^{++}}'$ in comparison to the other states depends on the model; in most models (Table\,\ref{tab:modelmasses}) the mass ordering is
\begin{align}
    M_0<M_1<M_2<M_0' \, ,\label{eq:ordering2}
\end{align}
and this is true of the model we use for our calculations, as shown generally in Ref.~\cite{Anwar:2023svj}. Some other models have a different ordering, such as $ M_0<M_1<M_0'<M_2$.

In our discussion on quark models we will assume the same assignment as is relevant to diquark models, namely $X(6400)$ and $X(6600)$ having $0^{++}$ and $2^{++}$ quantum numbers, respectively. This is partly to facilitate a comparison with diquark models, but also because the corresponding mass splitting is consistent with the predictions of a simple model whose parameters are fit to conventional mesons. The assignment is also qualitatively consistent with the experimental observation that the peak associated with $X(6400)$ is less prominent compared to $X(6600)$, as we argue later in the paper.

\section{Spectroscopy}
\label{sec:spectrum}

On general grounds, we expect the dynamics of $cc\c\c$ states to be described by pair-wise interactions between quark constituents, as distinct from (for example) molecular degrees of freedom (interacting colour-singlet quarkonia~\cite{Brambilla:2015rqa,Dong:2020nwy,Albuquerque:2020hio,Dong:2021lkh,Niu:2022jqp}) or effective diquarks. This is because the characteristic distance scale of an all-heavy tetraquark $QQ \bar Q \bar Q$, with quark mass $m_Q$, is of the order $1/(m_Q \alpha_s) \sim 1/(m_Q v)$, where $\alpha_s$ is the strong coupling constant and $v$ is the quark velocity. In this case, the dynamics of the system are expected to be dominated by the short-distance OGE interaction and the potential can be treated as pair-wise, quark-level interactions.

In Ref.~\cite{Anwar:2023svj} we compared a number of different models for tetraquark states, differing according to whether quarks or diquarks are the relevant degrees of freedom, and whether the constituents have effective masses, or instead dynamical masses which are treated in the Schr\"odinger equation. Our findings are that for S-wave states with either one or two quark flavours, we may characterise the spectrum for all models within the framework of the chromomagnetic quark model, with Hamiltonian
\be
H=\M
- \sum_{i<j} C_{ij}~\l_i\cdot\l_j~\s_i \cdot\s_j\,,\label{eq:h:chromo}
\ee
where $\M$ is the centre of mass, $\l_i$ and $\s_i$ are the $SU(3)$ colour and $SU(2)$ spin (Pauli) matrices of quark $i$, and $C_{ij}$ are (positive) parameters which depend on quark flavours. The spectrum applicable to quark models comes from diagonalising $H$ in the full basis of states $\f_2$, $\f_1$, $\f_0$ and $\f_0'$; the two scalar states are orthogonal combinations of $\f_0$ and $\f_0'$, with mixing due to the $\l_i\cdot\l_j~\s_i \cdot\s_j$ term. The spectrum of diquark models \cite{Maiani:2004vq,Maiani:2005pe,Drenska:2008gr,Drenska:2009cd,Ali:2009pi,Ali:2019npk,Ali:2019roi}, on the other hand, can be obtained from the same Hamiltonian, but instead using a truncated basis of wavefunctions with only $\f_2$, $\f_1$ and $\f_0$, but not $\f_0'$.

In the chromomagnetic model (and similarly in the simplest diquark model) the parameters $\M$ and $C_{ij}$ are essentially phenomenological. Typically $\M$ is taken as the sum of quark (or diquark) masses, with constraints derived from masses of mesons and baryons. The couplings $C_{ij}$ are assumed to scale inversely with quark masses, and can also be fit to mesons and baryons; see for example Refs.~\cite{Buccella:2006fn,Deng:2020iqw,Weng:2020jao}.

However these parameters can also be interpreted in the framework of dynamical models, where  quarks (or diquarks) are treated in the Schr\"odinger equation. In Ref.~\cite{Anwar:2023svj} we showed that the non-relativistic quark potential model reduces to the chromomagnetic model, in a symmetry limit where the spatial wavefunction of a $c\c$ pair within the tetraquark is the same as that of a $cc$ or $\c\c$ pair, and where the spin-dependent (chromomagnetic) interactions are treated in perturbation theory. In this comparison, $\M$ is the eigenvalue of the unperturbed Hamiltonian, and so should be understood as absorbing not only the quark rest masses, but also their kinetic energy, as well as the effects of the QCD confining interaction. In the same comparison (see also Ref.~\cite{Capstick:2000qj}) the coefficients $C_{ij}$ are
\begin{align}
       C_{ij}= \frac{\pi}{6}\, \frac{ \alpha_s}{m^2} \, \<\delta^3 (\mathbf r_{ij})\>\,,
\end{align}
where $\alpha_s$ is the (effective) strong coupling constant of QCD, $m$ is the quark mass, and the delta function in the relative quark coordinates $\mathbf r_{ij}$ is integrated over the spatial wavefunctions. 

In a similar way, the parameters $\M$ and $C_{ij}$ of the chromomagnetic model Hamiltonian can also be interpreted within the framework of diquark potential models, in which the hyperfine splitting is associated with effective diquark spin operators. Again, the correspondence applies when $H$ is evaluated in the truncated colour basis.

Regardless of whether the degrees of freedom are quarks or diquarks, and whether their masses are effective or dynamical, when applying the Hamiltonian \eqref{eq:h:chromo} to $cc\c\c$ systems, there are only two independent couplings 
\begin{align}
	C_{cc}&=C_{12}=C_{34}\label{eq:C1}\,,\\
	C_{c\c}&=C_{13}=C_{14}=C_{23}=C_{24}\,,\label{eq:C2}
\end{align}
and it is convenient to express the mass spectrum in terms of their ratio,
\begin{align}
    R=\frac{C_{c\c}}{C_{cc}} \,. \label{R-ratio}
\end{align}
For many of our calculations, we will assume $R=1$ which, in the quark potential model, is equivalent to assuming  that the spatial wavefunctions of $c\c$ pairs are identical to those of $cc$ and $\c\c$ pairs, as in for example Refs.~\cite{Anwar:2017toa,Zhang:2022qtp,Anwar:2023svj}.

In Ref.~\cite{Anwar:2023svj} we derived the mass spectrum of the Hamiltonian \eqref{eq:h:chromo}. In the quark model, the masses of the scalar ($M_0$, $M_0'$), axial ($M_1$) and tensor ($M_2$) states are, in increasing mass order,
\begin{align}
    M_0&=\M+\frac{4}{3}C_{cc}\left(5-4R-\Delta\right),\label{eq:m0}\\
    M_1&=\M+\frac{16}{3}C_{cc}(1-R),\label{eq:m1}\\
    M_2&=\M+\frac{16}{3}C_{cc}(1+R),\label{eq:m2}\\
    M_0'&=\M+\frac{4}{3}C_{cc}\left(5-4R+\Delta\right),\label{eq:m0p}
\end{align}
where
\begin{align} 
    \Delta=\sqrt{232R^2+8R+1}\, \label{Delta}.
\end{align}
In the diquark model, the axial ($M_1$) and tensor ($M_2$) are as above, but in place of $M_0$ and $M_0'$ there is only scalar state, with
\begin{align}
M_0&=\M+\frac{16}{3}C_{cc}(1-2R) \,. \label{eq:mf0} 
\end{align}

Naively we may expect that diquarks are a useful concept if $c\c$ interactions are small compared to $cc$ and $\c\c$ interactions, namely for small $R$. It is therefore interesting to note~\cite{Anwar:2023svj} that if we take the small $R$ limit of the chromomagnetic model, the masses $M_0$, $M_1$ and $M_2$ are identical to the corresponding masses in the diquark model; here we are using the approximation $\Delta\approx 1+4R$, which is suitable for small $R$. In this sense we can regard the diquark model as the small $R$ limit of the quark model, except for the missing heavier scalar ($M_0'$) which, in diquark models, is absent by construction. The small $R$ limit (namely $C_{cc}\gg C_{c\c}$) can be regarded as considering the dominant spin interactions to be those within each diquark, whereas spin interactions between quarks in different diquarks are suppressed, as in for example Ref.~\cite{Maiani:2014aja}.

\section{Interpretation of LHC States}
\label{sec:interpretation}

Let us now see how the predicted spectra compare to experimental data. We work initially in the symmetry limit ($R=1$), and for the parameter
\begin{align}
    C\equiv C_{c\c}=C_{cc}\,,
\end{align}
we adopt $C=5.0\pm 0.5$~MeV, on the basis of previous fits to meson and baryon spectra~\cite{Buccella:2006fn,Liu:2019zoy,Deng:2020iqw}. Using this value, we may estimate the mass splittings in the multiplet using the equations \eqref{eq:m0}-\eqref{eq:mf0}. To compare with experimental data, we are particularly interested of course in the splittings among the states which could in principle be visible in the $\jp\jp$ spectrum. In the diquark model, there are two such states ($0^{++}$ and $2^{++}$), and their splitting
\begin{align}
    M_2-M_0=16C= 80\pm 8~\textrm{MeV}\label{eq:splitting2}
\end{align}
is too small to match any pair of states measured in experimental data (see Table\,\ref{tab:expdata}). On the other hand, in the quark model, there are three possible states ($0^{++}$, $2^{++}$, ${0^{++}}'$), and with the same coupling the splittings are considerably larger. In particular, we notice that the splitting between the lower two
\begin{align}
    M_2-M_0=\frac{4}{3}(7+\sqrt{241})C= 150\pm 15~\textrm{MeV}\label{eq:splitting1}
\end{align}
is very close to the experimental splitting between $X(6400)$ and $X(6600)$. (Note that the central value of the mass of $X(6600)$ at CMS is somewhat lower than its name suggests: see Table\,\ref{tab:expdata}.) This motivates our preferred assignment of $X(6400)$ and $X(6600)$ as the scalar and tensor $cc\c\c$ states, respectively. This assignment is further supported by the strong decay patterns, which will be discussed in Sec.\,\ref{sec:decays}.

An important caveat here is that the ``state'' we are referring to as $X(6400)$ is not claimed as such by ATLAS, though it is clearly visible in their data, and they provide measured parameters (see Table\,\ref{tab:expdata}). The state is not reported by CMS, although there are hints in their spectrum for some enhancement in the same mass region.

In comparison to $X(6400)$, the state $X(6600)$ is more well-established, having been observed and measured at both CMS and ATLAS (with consistent parameters). For this reason, we fix the parameters of our model to $X(6600)$, using the (more precise) mass from CMS~\cite{CMS:2023owd}. Considering this assignment as an input to the chromomagnetic model, and fixing $R=1$, the central mass $\M$ can be extracted for different values of $C$, which further can be used to predict the masses of the other members of $S$-wave multiplet.

\begin{table}
\centering
\begin{tabular}{ll}
\hline\hline
$J^{PC}$\qquad\qquad & Mass (MeV) \\ [0.5ex] 
\hline
$0^{++}$ & 6402 $\pm$ 15\\
$1^{+-}$& 6499 $\pm$ 11  \\
$2^{++}$ & 6552 $\pm$ 10\quad(Input)\\
 $0^{++'}$ & 6609 $\pm$ 16  \\ [1ex]
\hline
\end{tabular}
\caption{Predicted spectrum of $S$-wave $cc \bar c \bar c$ states in the quark model, having fixed the tensor ($2^{++}$) mass to the CMS value~\cite{CMS:2023owd} for $X(6600)$, and using equations \eqref{eq:m0}-\eqref{eq:m0p}, with $C=5.0\pm 0.5$~MeV and $R=1$.}
\label{table:masses}
\end{table}

Adopting the preferred value of $C=5.0\pm 0.5$~MeV, our predictions for the masses of lowest-scalar $0^{++}$, axial-vector $1^{+-}$, and higher scalar $0^{++'}$ are given in Table\,\ref{table:masses}, where the uncertainties are due to the experimental uncertainty in $M_2$ and the quoted uncertainty in $C$. The lowest scalar is of considerable interest: our prediction for its mass is $M_0=6402 \pm 15$~MeV, which is consistent with the $X(6400)$ enhancement at ATLAS. Our predictions for the other two states can be tested in various decay channels, and we return to this point in Section\,\ref{sec:decays}.

\begin{figure}
   \centering
    \includegraphics[width=0.8\textwidth]{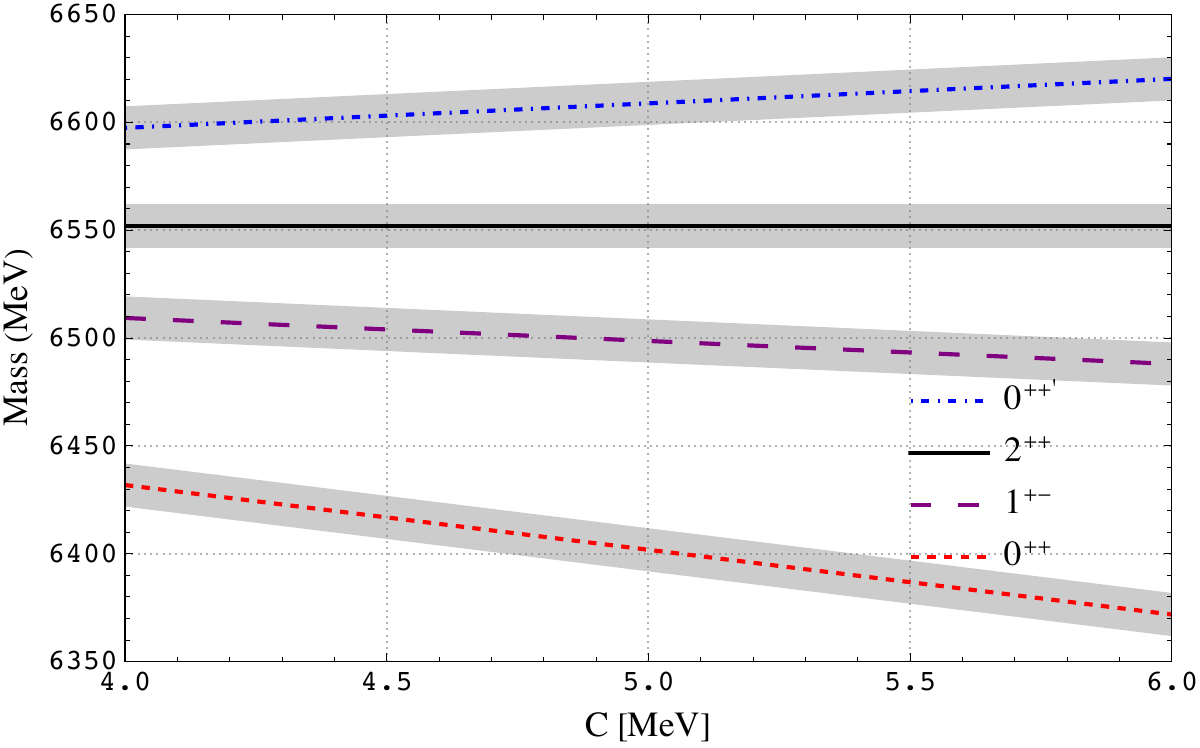}
    \caption{Masses of $S$-wave $cc \bar c \bar c$ states in the quark model, as a function of coupling strength $C$, where the  tensor state~($2^{++}$) is fixed to the $X(6600)$ measured at CMS~\cite{CMS:2023owd}, and masses of the remaining states are computed from equations~\eqref{eq:m0}-\eqref{eq:m0p}, with $R=1$.}
    \label{massplot}
\end{figure}

To illustrate the sensitivity of our results to $C$, we show in Fig.\,\ref{massplot} the predicted masses of the multiplet as a function of $C$, where the error bands are due to the experimental uncertainty in the input mass of the $2^{++}$ state. The message of this plot is that the predictions are quite robust. The mass of the lighter scalar ($0^{++}$) is rather sensitive to $C$, but over the full range of $C$ shown in the plot, it remains consistent with the ATLAS mass for $X(6400)$, within errors. The masses of the axial ($1^{+-}$) and heavy scalar (${0^{++}}'$) are much less sensitive to  $C$, with a fairly small variation across the full range of $C$ shown in the plot. 

In determining a suitable range of $C$, we have been guided so far by fits (such as Refs.~\cite{Buccella:2006fn,Liu:2019zoy,Deng:2020iqw}) to the spectrum of conventional hadrons. Of course one may question the validity of this approach, noting that there is no symmetry principle which equates the strength of colour-magnetic interactions inside a tetraquark to those in conventional mesons or baryons. 

Hence as a check on our conclusions, we now consider an alternative approach, extracting the model parameters directly from the tetraquark mass spectrum, rather than the spectra of conventional hadrons. Thus instead of taking $X(6600)$ and $C$ as inputs, and predicting $X(6400)$, we take the masses of $X(6600)$ and $X(6400)$ as inputs, and extract  the implied value of $C$. For $X(6400)$ we use the ATLAS~\cite{ATLAS:2023bft} mass (see Table\,\ref{tab:expdata}), since only ATLAS has measured parameters for this state. For $X(6600)$ we again take the CMS value~\cite{CMS:2023owd}, due to its higher precision compared to the other experiments.
As before, we assign $X(6600)$ and $X(6400)$ as the $2^{++}$ and $0^{++}$ states, respectively. The fitted value of coupling strength in the chromomagnetic model is then $C = 4.7 \pm 2.9$ MeV, where the large uncertainty is dominated by the input mass of $X(6400)$. This is in good agreement with the value $C=5.0\pm 0.5$~MeV extracted from the meson spectrum, which supports the validity of assuming a common coupling strength in both tetraquarks and conventional hadrons.

So far we have assumed equal couplings for $cc$ and $c\c$ interactions ($R=1$), which takes no account of the spatial variation in the $c\c$ wavefunctions compared to $cc$ (and $\c\c$). In order to generalise our results somewhat, we now relax this assumption, and allow for $C_{c\c}\ne C_{cc}$, namely $R\ne 1$. We will also no longer require that the values of these couplings are constrained by comparison the spectra of conventional hadrons; instead, we will assume that they can be adjusted to reproduce the masses of  $X(6400)$ and $X(6600)$ as the scalar and tensor states, respectively. In this case the diquark model, which had previously been ruled out on the basis of the mass splitting, becomes a possibility.

The splitting in diquark models is sensitive to 
$C_{c\c}$ (not $C_{cc}$), specifically
 \begin{align}
     M_2-M_0=16C_{c\c}\,.
 \end{align}
To accommodate the (approximately) 150~MeV splitting between $X(6400)$ and $X(6600)$ implies $C_{c\c}\approx 9.4$~MeV, somewhat larger than the value indicated by the meson and tetraquark spectrum. 

In quark models, on the other hand, the splitting is a function of both $C_{cc}$ and $C_{c\c}$ -- or equivalently $C_{cc}$ and the ratio $R=C_{c\c}/C_{cc}$,
 \begin{align}
     M_2-M_0=\frac{4}{3}C_{cc}\left(8R-1+\Delta\right).\label{eq:split02}
 \end{align}
We already know that the combination $C_{cc}=5 \pm 0.5$~MeV and $R=1$ generates the required 150~MeV splitting, but clearly these parameters are not unique, so it is interesting to explore how our predictions depend on these parameters.

Having assigned $X(6400)$ and $X(6600)$ as the scalar and tensor states, respectively, 
we may then predict the masses of the additional partner states, using the relations derived in Ref.~\cite{Anwar:2023svj}, and which also follow straightforwardly from equations \eqref{eq:m0}-\eqref{eq:mf0}. These predictions offer a key experimental test to distinguish models. In diquark models, there is just one further state in the multiplet (the axial) with mass
\begin{align}
M_1=\frac{1}{3}\left(2M_0+M_2\right).\label{eq:diq:relation}
\end{align}
In quark models, by contrast, there are two further states (axial and scalar), whose masses depend on $R$,
\begin{align}
    M_1&=M_0+\frac{\Delta-1}{\Delta-1+8R}(M_2-M_0),\label{eq:relation1}\\
    M_0'&=M_0+\frac{2\Delta}{\Delta-1+8R}(M_2-M_0).\label{eq:relation2}
\end{align}

\begin{figure}
   \centering
    \includegraphics[width=\textwidth]{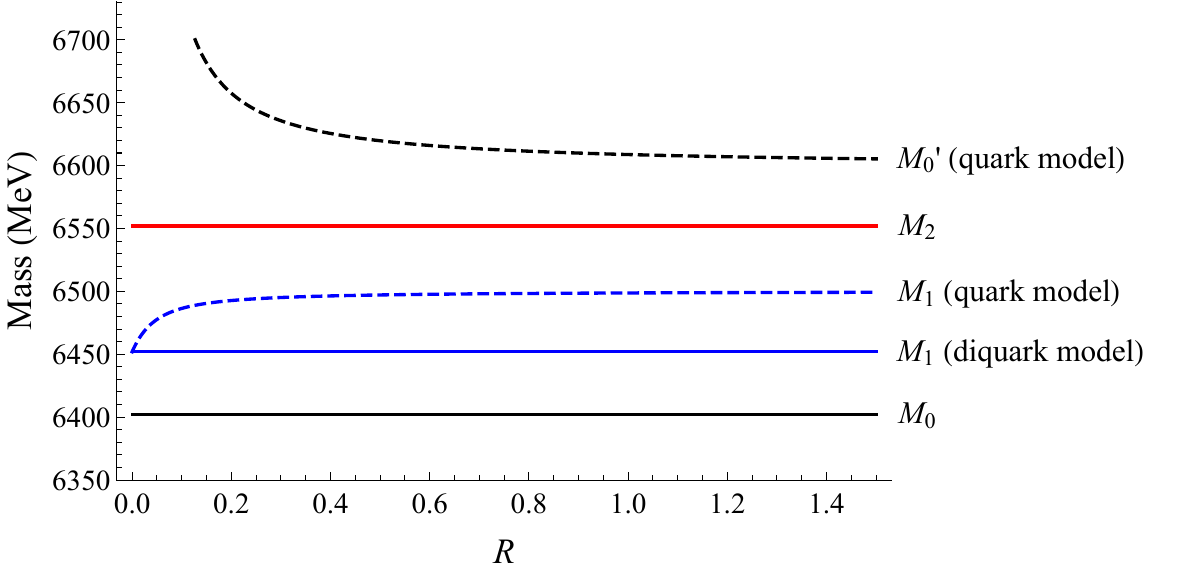}
    \caption{The spectrum of states where $M_0$ and $M_2$ are fixed to the masses of $X(6400)$ and $X(6600)$ as in Table~\ref{tab:modelmasses},  and the remaining masses are predictions from the mass relations \eqref{eq:diq:relation}-\eqref{eq:relation2}.}
    \label{fig:spectrum}
\end{figure}

In Fig.\,\ref{fig:spectrum} we show these predictions as a function of $R$ where, for the sake of comparison with our previous results, we have fixed $M_0$  and $M_2$ to the values in Table\,\ref{tab:modelmasses}. The mass of the axial state $M_1$ differs for quark models and diquark models, and the heavier scalar $M_0'$  is of course a feature of the quark model only.

An interesting feature of Fig.\,\ref{fig:spectrum} is that the predicted masses of the axial $M_1$ in quark and diquark models become degenerate in the limit $R\to 0$, a result which we proved in Ref.~\cite{Anwar:2023svj}. However this limit is not physical once we have fixed $M_2-M_0=150$~MeV, since for small $R$ we have $\Delta\sim 1+4R$ which implies, from equation \eqref{eq:split02}, that  $C_{cc}$ blows up. To avoid this unphysical situation, we focus on values of $R$ which are not close to zero, and it is reassuring that in this region our quark model predictions for $M_1$ and $M_0'$ are quite insensitive to $R$. It suggests that the values quoted in Table\,\ref{tab:modelmasses} (corresponding to $R=1$) are quite reliable.

In the same region ($R$ not close to zero), the predictions for the axial mass $M_1$ in quark and diquark models are very different, which offers a key experimental test of models. The quark model prediction is weakly dependent on $R$; the value at $R=1$ is, from Table\,\ref{tab:modelmasses}, $M_1=6499$~MeV. For comparison the diquark model result, from equation~\eqref{eq:diq:relation}, is $M_1=6452$~MeV, independently of $R$.

Another way of phrasing the results is in terms of the ratio  $\Delta_2/\Delta_1$ of splittings 
\begin{align}
\Delta_1&= M_1 -M_0\label{eq:diq:m0}\, ,\\
\Delta_2&= M_2 -M_1\label{eq:diq:m1}\, . 
\end{align}

In diquark models, from equation~\eqref{eq:diq:relation}, we expect $\Delta_2/\Delta_1=2$. The result is exact for diquark models with effective masses, and in diquark potential models in which spin-spin interactions are treated perturbatively. For potential models not relying on perturbation theory, the relation is satisfied approximately, $\Delta_2/\Delta_1 \approx 2$, becoming closer to exact for $bb\b\b$ states~\cite{Annathesis}, where the spin splittings are smaller, and perturbation theory is more reliable.

\begin{figure}
   \centering
    \includegraphics[width=0.8\textwidth]{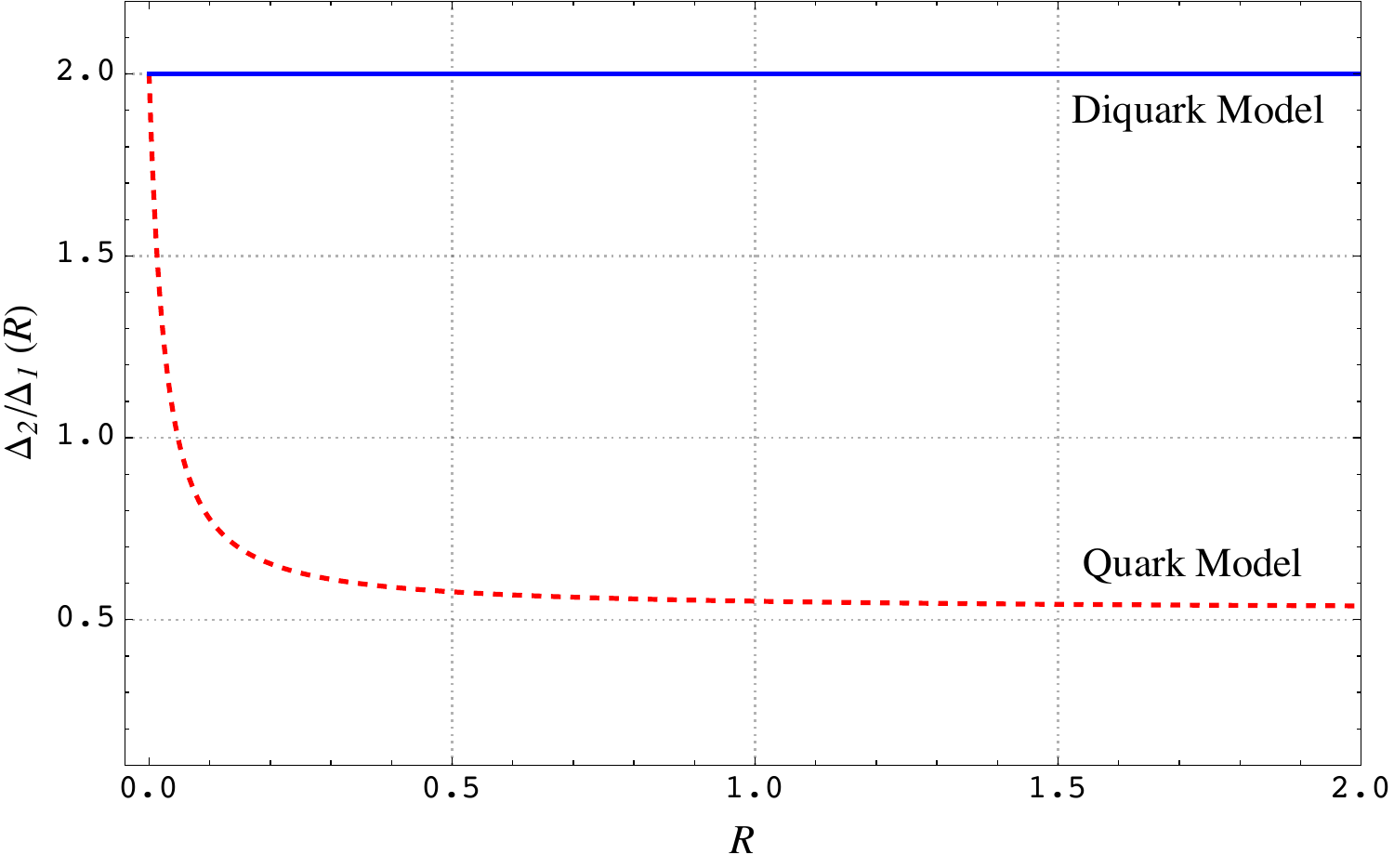}
    \caption{The ratio  $\Delta_2 /\Delta_1$ of the mass splittings defined in equations~\eqref{eq:diq:m0}-\eqref{eq:diq:m1}, as a function of $R$, in the quark model (red curve) and diquark model (blue line). Notice that the models agree in the (unphysical) limit $R\to 0$, as described in the text.}
    \label{Rplot}
\end{figure}

By contrast, the quark model prediction  for the ratio $\Delta_2/\Delta_1$  is very different, and this offers a key experimental test of models. From equations~\eqref{eq:relation1}-\eqref{eq:relation2}, with $R=1$ the quark model ratio is $\Delta_2/\Delta_1=0.55$. Notably, the dependence of the ratio on $R$ is rather weak, in the physically relevant region of $R$ not close to zero. In Figure.\,\ref{Rplot} we show the ratio $\Delta_2/\Delta_1$ in the quark model as a function of $R$, noting in particular that as $R\to 0$ we recover the diquark model result $\Delta_2/\Delta_1=2$.  For a reasonable range of $R$ (not close to zero) the ratio is well separated from 2; an experimental spectrum with this pattern would indicate that quarks (not diquarks) are the relevant degrees of freedom.

\section{Decays}
\label{sec:decays}

The other main focus of this study is the strong decay patterns of all-charm tetraquarks. Absolute predictions for strong decays involve matrix elements integrated over hadronic wavefunctions, which are very much model-dependent. To get more robust predictions, here we concentrate on relations among strong decays, by comparing transitions which share (approximately) the same spatial matrix element, but which different in their colour and spin matrix elements.

\subsection{Overview}
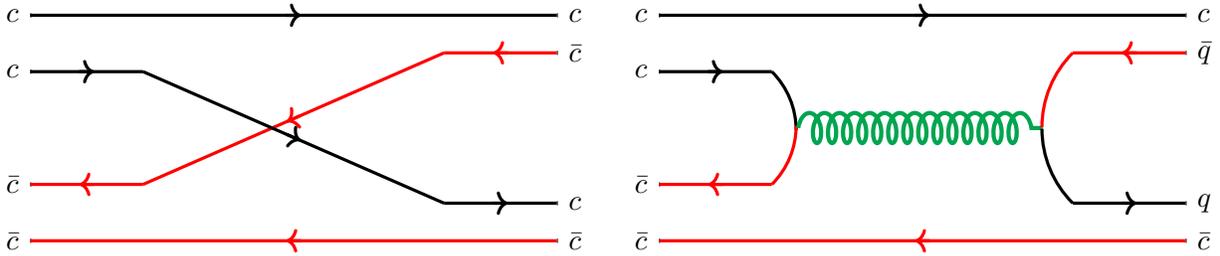
\begin{figure}
   \centering
\begin{tikzpicture}
  \draw[ultra thick] (-3.5,-0.5) node[left] {\textbf{$\bar c$}} -- (-3.5,-0.5) node {};    
  \draw[ultra thick] (3.5,-0.5) node[right] {\textbf{$\bar c$}} -- (3.5,-0.5) node {};     
  \draw[ultra thick] (-3.5,0.25) node[left] {\textbf{$\bar c$}} -- (-3.5,0.25) node {};        
  \draw[ultra thick] (3.5,0) node[right] {\textbf{$c$}} -- (3.5,-0) node {};         
  \draw[ultra thick] (-3.5,2.5) node[left] {\textbf{$c$}} -- (-3.5,2.5) node {};    
  \draw[ultra thick] (3.5,2.5) node[right] {\textbf{$c$}} -- (3.5,2.5) node {};     
  \draw[ultra thick] (-3.5,1.75) node[left] {\textbf{$c$}} -- (-3.5,1.75) node {};        
  \draw[ultra thick] (3.5,2) node[right] {\textbf{$\bar c$}} -- (3.5,2) node {};         
  \begin{scope}[very thick, every node/.style={sloped,allow upside down}]
     \draw [red] (3.5,-0.5) --node {\midarrow} (-3.5,-0.5) ;
    \draw [red] (-2,0.25)  --node {\midarrow} (-3.5,0.25) ;
    \draw [red,very thick] (2,2) -- node {\midarrow} (-2,0.25) ;
    \draw [red] (3.5,2)  --node {\midarrow} (2,2) ;
    
    \draw (-3.5,2.5) --node {\midarrow} (3.5,2.5);
    \draw (-3.5,1.75) --node {\midarrow}  (-2.,1.75);
    \draw [very thick] (-2.,1.75) -- node {\midarrow} (2.,0);
    \draw (2.,0) --node {\midarrow} (3.5,0);
  \end{scope}
\end{tikzpicture}
\hspace{0.2cm}
\begin{tikzpicture}

  \draw[ultra thick] (-3.5,-0.5) node[left] {\textbf{$\bar c$}} -- (-3.5,-0.5) node {};    
  \draw[ultra thick] (3.5,-0.5) node[right] {\textbf{$\bar c$}} -- (3.5,-0.5) node {};     
  \draw[ultra thick] (-3.5,0.25) node[left] {\textbf{$\bar c$}} -- (-3.5,0.25) node {};        
  \draw[ultra thick] (3.5,0) node[right] {\textbf{$q$}} -- (3.5,-0) node {};         

  \draw[ultra thick] (-3.5,2.5) node[left] {\textbf{$c$}} -- (-3.5,2.5) node {};    
  \draw[ultra thick] (3.5,2.5) node[right] {\textbf{$c$}} -- (3.5,2.5) node {};     
  \draw[ultra thick] (-3.5,1.75) node[left] {\textbf{$c$}} -- (-3.5,1.75) node {};        
  \draw[ultra thick] (3.5,2) node[right] {\textbf{$\bar q$}} -- (3.5,2) node {};         
%
  \begin{scope}[very thick, every node/.style={sloped,allow upside down}]
    
    \draw [red] (3.5,-0.5) --node {\midarrow} (-3.5,-0.5) ;
    \draw [red] (-2,0.25)  --node {\midarrow} (-3.5,0.25) ;
    \draw [red] (3.5,2)  --node {\midarrow} (2,2) ;
  \draw       (-2,1.75) arc(45:-0:1.1);
  \draw [red] (-2,.25) arc(-45:-0:1.1);
  
    \draw[ultra thick,Green, decorate,decoration={coil,aspect=0.5,segment length=2mm,amplitude=2mm}]  (-1.65,1.) -- (1.6,1) {}; 
    \draw [red]      (2,2) arc(-225:-180:1.4);
  \draw   (2,0) arc(225:180:1.4);  
    \draw (-3.5,2.5) --node {\midarrow} (3.5,2.5);
    \draw (-3.5,1.75) --node {\midarrow}  (-2.,1.75);
    \draw (2.,0) --node {\midarrow} (3.5,0);
  \end{scope} 
\end{tikzpicture}
\caption{Quark rearrangement (left) and annihilation (right) decays of $cc\c\c$ tetraquarks.}
    \label{decaysfig}
\end{figure}

The two main decay processes we will consider are shown in Fig.\,\ref{decaysfig}. As the states are above $\jp\jp$ threshold, their dominant decay is expected to be via  a quark rearrangement process (we refer to this as \emph{rearrangement decays}), where the $cc\c\c$ state dissociates into combinations of $J/\psi$ or $\eta_c$ mesons (depending on quantum numbers), as shown in the left panel of Fig.\,\ref{decaysfig}. The discovery mode $\jp\jp$ is of course an example of such a process.

Another possibility is that the $cc\c\c$ state decays into \dordstar/ via annihilation of a spin-1 colour-octet $c\c$ pair into a gluon,\footnote{In fact, this decay mode is expected to be the dominant decay for $cc\c\c$ states below the threshold of $2J/\psi$~\cite{Anwar:2017toa}.} namely $c \bar c \to g \to q \bar q$, as shown in the right panel of Fig.\,\ref{decaysfig} (we refer to these as \emph{annihilation decays}). Relative to rearrangement decays, these have a larger phase space, but are suppressed due to having two vertices of the strong interaction (albeit, a weaker suppression than the annihilation of a $J/\psi$ into light hadrons, which involves three gluons). These channels are of particular interest because, as mentioned previously, they can be studied in  future experiments such as STCF.

For both processes (rearrangement decays and annihilation decays), the relative strengths of decays for different initial or final states are sensitive to the colour-spin wavefunctions, which are defined in terms of the basis states $\f_2$, $\f_1$, $\f_0$ and $\f_0'$ in equations \eqref{eq:basis2}-\eqref{eq:basis0p}. For the tensor and axial states, the colour-spin wavefunctions  are  $\f_2$ and $\f_1$, regardless of the model. For the scalar states, however, the wavefunctions differ according to the model. In diquark models, there is a single scalar state $\f_0$, corresponding to the pure ``hidden'' colour triplet configuration. In quark models, there are two scalars, which are admixtures of the colour triplet and colour sextet configurations:
\begin{align}
\begin{pmatrix}\|0^{++}\>\\\|0^{++\prime}\>\end{pmatrix}&=
\begin{pmatrix}\cos\theta&\sin\theta\\-\sin\theta&\cos\theta\end{pmatrix}\begin{pmatrix}\|\f_0\>\\ \|\f_0'\>\end{pmatrix} \label{eq:scalars}\ .
\end{align}

To get results which are applicable to both quark and diquark models, we will evaluate relative partial widths as a function of the mixing angle $\theta$. Predictions for the diquark model then follow by fixing $\theta=0$ and evaluating the partial widths for the state $0^{++}$, ignoring the other scalar ${0^{++}}'$, which is absent by construction. For the quark model, instead, we include all states in the spectrum, and allow $\theta$ to vary. In Ref.~\cite{Anwar:2023svj} we derived an expression for the mixing angle 
\begin{align} \label{angle}
    \theta= \tan^{-1}\left(\frac{\Delta-1-4R}{6\sqrt 6 R}\right),
\end{align}
where $R$ and $\Delta$ are given by equations \eqref{R-ratio} and \eqref{Delta}, respectively. The result applies to the chromomagnetic quark model, and also to quark potential models in perturbation theory,  subject to the additional symmetry constraint discuss previously (identical spatial wavefunctions for $cc$ and $c\c$ pairs). In both cases it is natural to adopt $R=1$, which implies $\theta=35.6\degree$, which is the angle we will use when quoting numerical predictions for the quark model.

\subsection{Quark Rearrangement Decays}
\label{sec:rearr}
The decay channels accessible in S-wave by quark rearrangement are restricted by charge conjugation symmetry, and the possibilities are summarised in equations~\eqref{eq:rearr1}-\eqref{eq:rearr3}. The interaction Hamiltonian for this transition does not involve any strong interaction vertex, hence is zeroth order in the strong coupling, $ \hat{H}_0 \sim \alpha_s^0$. 

There are two possible decay topologies, distinguished according to which $c$ quark is paired with which $\c$ after quark rearrangement. Careful evaluation of these diagrams shows that they provide exactly the same contribution. However we suppress the overall factor of 2, which is common to all transitions and so cancels when comparing decay rates. 

The specific diagram we calculate is that shown in the left panel of Fig.\,\ref{decaysfig}. The transition amplitude factorises into spin, colour, and spatial parts. Taking $0^{++}\to \eta_c\eta_c$ as an example, we have
\be \label{amp1}
\langle \eta_c\eta_c| \hat{H}_0 | 0^{++} \rangle =  \phi_{\rm spin} \times\phi_{\rm colour} \times A(p),
\ee 
where $\phi_{\rm spin}$ and $\phi_{\rm colour}$ are matrix elements of the spin and colour wavefunctions, and $A(p)$ is the spatial part, which depends on the hadron spatial wavefunctions and the decay momentum $p$.

We will assume that the operator $\hat H_0$ itself is independent of spin and colour, in which case the corresponding matrix elements $\phi_{\rm spin}$ and $\phi_{\rm colour}$  are obtained via Fierz rearrangement.
For the topology in Fig.\,\ref{decaysfig} (left), the matrix element $\phi_{\rm spin}$  is the coefficient in the recoupling of the spin wavefunctions,
\begin{subequations}
\label{fierz}
\begin{align}
\|\{(cc)^1 (\c \c)^1 \}^2\> &= \|\{(c \c)^1 (c \c)^1\}^2 \>, \label{fierz4}\\
\|\{(cc)^1 (\c \c)^1\}^1 \> &= \frac{1}{\sqrt{2}} \|\{(c \c)^0 (c \c)^1\}^1 \> + \frac{1}{\sqrt{2}} \|\{(c \c)^1 (c \c)^0 \}^1\>,\label{fierz3}\\
\|\{(cc)^1 (\c \c)^1\}^0 \>&= \frac{\sqrt{3}}{2} \|\{(c \c)^0 (c \c)^0\}^0 \> - \frac{1}{2} \|\{(c \c)^1 (c \c)^1\}^0 \>, \label{fierz1}\\
\|\{(cc)^0 (\c \c)^0\}^0 \>&=\frac{1}{2} \|\{(c \c)^0 (c \c)^0\}^0 \> + \frac{\sqrt{3}}{2} \|\{(c \c)^1 (c \c)^1 \}^0\>,\label{fierz2}
\end{align}
\end{subequations}
while $\phi_{\rm colour}$ is the coefficient in the colour recoupling
\begin{subequations}
\label{colour}
\begin{align}
|(cc)_{\bar\three}(\c\c)_\three \> & = \sqrt{\frac13} | (c\c)_{\one}(c\c)_\one \> - \sqrt{\frac23} | (c\c)_{\eight}(c\c)_\eight \>, \label{colour1} \\
|(cc)_\six(\c\c)_{\bar\six} \> & =  \sqrt{\frac23} | (c\c)_{\one}(c\c)_\one \>  + \sqrt{\frac13} | (c\c)_{\eight}(c\c)_\eight \> \,. \label{colour2}
\end{align}
\end{subequations}
In this way we obtain, for example
\begin{align}
        \<\eta_c\eta_c\|\hat H_0\|0^{++}\>&= \left(\frac{\cos\theta}{2}+\frac{\sin\theta}{\sqrt 6}\right)A(p)\,.\label{eq:decayexample}
\end{align}
The amplitudes for all other transitions, obtained in the same way, are in the Appendix.

The spatial part of the transition amplitude $A(p)$, which is a function of the decay momentum $p$, could be obtained by integrating $\hat H_0$ over the spatial wavefunctions of the hadrons involved. This is of course model-dependent, and difficult to calculate reliably. However when comparing related transitions (such as $0^{++}\to\eta_c\eta_c$ and $2^{++}\to\jp\jp$) we may assume that the spatial part is the same, which is valid to the extent that the decay momenta are similar (noting that for S-wave transitions, $A(p)$ depends weakly on $p$), and assuming the same spatial wavefunctions for $0^{++}$ and $2^{++}$,  and for  $\eta_c$ and $\jp$. In this case, when comparing related transitions, the spatial part cancels, and the relative decay partial widths are controlled by $\phi_{\rm spin}$ and $\phi_{\rm colour}$. As an example, from the expressions in the Appendix we find
\begin{align}
    \frac{\Gamma(0^{++}\to\eta_c\eta_c)}{{\Gamma(2^{++}\to\jp\jp)}}=
    \frac{\omega(0^{++}\to\eta_c\eta_c)}{\omega(2^{++}\to\jp\jp)}\frac{1}{4}\left({\sqrt 3}\cos\theta+{\sqrt 2}\sin\theta\right)^2
      \end{align}
where $\omega$ is the phase space factor appropriate to each decay.

We will normalise all decay channels, as in this example, against the $2^{++}\to\jp\jp$ decay. This is partly because it is the only  $\jp\jp$ decay which does not depend on the mixing angle, and also because, in our preferred assignment, it corresponds to the prominent $X(6600)$ peak in $\jp\jp$, and thus offers a natural benchmark against which to measure other decay channels.

\begin{figure}[h!]
   \centering
    \includegraphics[width=0.8\textwidth]{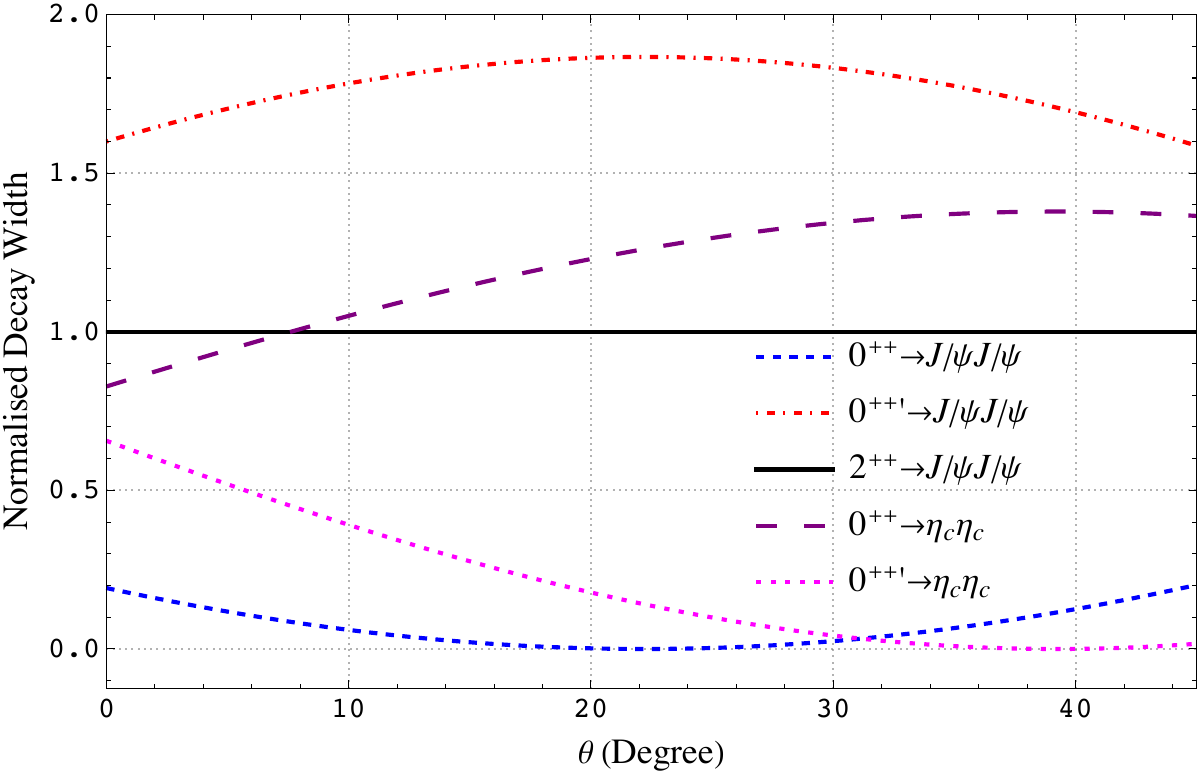}
    \caption{The ratio $\Gamma(X\to AB)/\Gamma(2^{++} \to J/\psi J/\psi)$ for different initial states $X$ and various hidden-charm final states $AB$, as a function of the scalar mixing angle $\theta$.}
    \label{mixing}
\end{figure}

Our results for the relative partial widths, normalised to $2^{++}\to\jp\jp$, are shown in Table\,\ref{table:decays} (for specific values of the mixing angle $\theta$) and Fig.\,\ref{mixing} (as a function of $\theta$). The phase space factors in each case have been computed using the masses from Table \ref{table:masses}. (We are ignoring the effect on the phase space factors of the variation of masses with mixing angle.)  

The natural mixing angle in the quark model, as discussed previously, is $\theta=35.6\degree$. However in Table\,\ref{table:decays} we also quote the results for $\theta=0\degree$, corresponding to no mixing. This is partly to give an indication of the pronounced effect of mixing on the relative partial widths. But also, as discussed previously, because it facilitates a comparison between quark and diquark models, where for the latter we take the $0^{++}$ entry with $\theta=0$, and ignore the $0^{++\prime}$ state, which is absent in the diquark model by construction. 

\begin{table}
 \renewcommand\arraystretch{1.2}
  \setlength{\tabcolsep}{10pt}
  \centering
\centering
\begin{tabular}{l|ll|ll|ll}
\hline\hline
\multirow{2}*{Final State}&\multicolumn{2}{c|}{$\theta=35.6\degree$} &\multicolumn{2}{c|}{$\theta=0\degree$} & \multirow{2}*{$2^{++}$} & \multirow{2}*{$1^{+-}$} \\
\cline{2-5}
& $0^{++}$ & $0^{++'}$ & $0^{++}$ & $0^{++'}$ &  &  \\ [1ex]
\hline
$J/\psi J/\psi$ & 0.072 & 1.76  & 0.19 & 1.60 & 1.0 & $-$\\
$\eta_c \eta_c$ & 1.38 & 0.01  & 0.83 & 0.66 & $\sim 0$ & $-$\\
$J/\psi \eta_c$ & $-$ & $-$  & $-$ &  $-$ & $-$ & 1.08 \\ [1ex]
\hline
\end{tabular}
\caption{The ratio $\Gamma(X\to AB)/\Gamma(2^{++} \to J/\psi J/\psi)$ for different initial states $X$ and various hidden-charm final states $AB$, computed as in equation~\eqref{eq:decayexample}.}
\label{table:decays}
\end{table}

A noteworthy feature of the predictions in Table\,\ref{table:decays} and Fig.\,\ref{mixing} is that the light scalar decay  $0^{++}\to J/\psi J/\psi$ is suppressed relative to the benchmark channel $2^{++}\to J/\psi J/\psi$. This applies regardless of mixing angle, although the suppression is stronger for quark model mixing compared to the no mixing case. Recalling our favoured scenario in which $X(6400)$ and $X(6600)$ are the $0^{++}$ and $2^{++}$ states, respectively, these predictions are qualitatively consistent with experimental data, in which the $X(6400)$ peak in $\jp\jp$ is less prominent that $X(6600)$ -- though of course the comparison takes no account of possible differences in the production cross section for the $0^{++}$ and $2^{++}$ states. 

Conversely, for the heavier scalar $0^{++'}$, which is expected in quark models but not diquark models, the decay $0^{++'}\to J/\psi J/\psi$ is enhanced relative to the benchmark channel $2^{++}\to J/\psi J/\psi$. Experimental search for structure in $\jp\jp$ spectrum near 6600~MeV (see Table~\ref{table:masses}) could therefore be quite revealing. Confirmation of a structure in this mass region would support the quark model scenario. Conversely, a lack of structure in this region would be less conclusive, as it could be that the heavier scalar $0^{++'}$ does not exist (as in the diquark model), or simply, that its production its suppressed.

Comparing the decays of the two scalars (Table\,\ref{table:decays} and Fig.\,\ref{mixing}) a distinctive feature is their relative rate into $\eta_c \eta_c$ and $J/\psi J/\psi$. In particular, the lighter scalar $0^{++}$ decays dominantly into $\eta_c \eta_c$, whereas the heavier scalar $0^{++'}$ decays dominantly to $J/\psi J/\psi$. This pattern applies regardless of mixing angle, although the relative size of $\eta_c \eta_c$ and $J/\psi J/\psi$ is sensitive to the mixing angle, and illustrates the importance of taking account of colour mixing, which is sometimes ignored. 
For example the dominant decay of the lighter scalar is enhanced by the mixing of different colour configurations, with the ratio $\Gamma(0^{++} \to \eta_c \eta_c)/\Gamma(2^{++}\to\jp\jp)$ increasing from 0.83 (no mixing) to 1.38 (quark model mixing). More dramatically, the equivalent ratio for the heavy scalar $\Gamma(0^{++'} \to \eta_c \eta_c)/\Gamma(2^{++}\to\jp\jp)$ decreases from 0.66 (no mixing) to just 0.01 (quark model mixing). 

Another way of phrasing these results is by a direct comparison of the two decay modes for each initial state. For the unmixed case (for $\theta=0\degree$) we have
\be
\frac{\Gamma(  0^{++} \to \eta_c \eta_c)}{\Gamma( 0^{++}\to J/\psi J/\psi) } = 4.29 \,,~~~~~
\frac{\Gamma(  0^{++'}\to\eta_c \eta_c)}{\Gamma( 0^{++'}\to J/\psi J/\psi) } = 0.41 \, , \label{scalardecays}
\ee
whereas for quark model mixing ($\theta=35.6\degree$) we have 
\be
\frac{\Gamma(  0^{++} \to \eta_c \eta_c)}{\Gamma( 0^{++}\to J/\psi J/\psi) } = 18.98 \,,~~~~~
\frac{\Gamma(  0^{++'}\to\eta_c \eta_c)}{\Gamma( 0^{++'}\to J/\psi J/\psi) } = 0.004 \, . \label{scalardecays2}
\ee
These results offer a simple test of our favoured scenario, in which the $X(6400)$ state is the light scalar $0^{++}$: we predict that it will decay prominently to $\eta_c\eta_c$ in comparison to $\jp\jp$. This applies to both diquark models and quark models, although the enhancement of $\eta_c\eta_c$ is significantly stronger in the latter case. We therefore urge an experimental study of the $\eta_c\eta_c$ spectrum, as a critical test of the existence of $X(6400)$ (which has not yet been confirmed at CMS), and to discriminate between quark and diquark models.

By contrast, in $\eta_c\eta_c$ decays we do not expect a signal for the heavier scalar $0^{++'}$. In quark models (with $\theta=35.6\degree$) the partial width is effectively zero (see above), while in diquark models the heavier scalar $0^{++'}$ is absent by construction.

To summarise our results for rearrangement decays, in the $J/\psi J/\psi$ spectrum there are currently two structures $X(6400)$ and $X(6600)$ which, in our approach, are $0^{++}$ and $2^{++}$ states.  A striking signature of the quark model (as compared to the diquark model) would be the discovery of a third structure ($0^{++'}$) in  $J/\psi J/\psi$, above $X(6600)$. The $\eta_c\eta_c$ spectrum has a characteristically different pattern; we predict a strong signal for $X(6400)$, but not for $X(6600)$ or the heavier scalar.

Finally we remark (see Table\,\ref{table:decays}) that the axial-vector $1^{+-}$ is expected to leave prominent signatures in $\eta_c J/\psi$ final state. Given that an initial search by Belle recently found an evidence for $e^+ e^- \to \eta_c J/\psi$ near the $\eta_c J/\psi$ threshold~\cite{Belle:2023gln}, studies with more data seems necessary and encouraging. This channel is of particular interest because, as discussed previously (see also Figs.\,\ref{fig:spectrum} and~\ref{Rplot}), the mass of the $1^{+-}$ state clearly discriminates between quark and diquark models.

\subsection{Annihilation Decays}
\label{sec:anni}
The dominant mechanism for the decay of a $cc\c\c$ state to open charm pairs such as \dordstar/ is illustrated in  the right panel of Fig.\,\ref{decaysfig}. As distinct from quark rearrangement decays, there are two strong interaction vertices, so the Hamiltonian is second order in $\sqrt{\alpha_s}$, namely $\hat{H}_2 \sim \alpha_s$. We make no attempt to compute absolute decay widths for such processes, which are necessarily highly model-dependent. Instead we follow a similar approach as in our discussion of quark rearrangement decays, and focus on relations among similar decays; these depend only on the spin and colour wavefunctions, and so can be more reliably calculated, and additionally, they offer more direct tests to distinguish between quark and diquark models.

The essential process is $(c \bar c)_{\eight}^1 \to g \to (q \bar q)_{\eight}^1$, where a spin-1 colour-octet $c\c$ pair annihilates, via a gluon, to a spin-1 colour-octet $q\q$ pair. There are four such diagrams, corresponding to which of the four possible $c\c$ pairs annihilate. Careful evaluation of these diagrams shows that they provide exactly the same contribution, so we will concentrate on just one of the four possible diagrams. We suppress the overall factor of 4 which would come from summing four equivalent diagrams, as this is common to all transitions, and so cancels when comparing decay rates.

Taking the $\f_J$ states of equations \eqref{eq:basis2}-\eqref{eq:basis0} as an example, the amplitude factorises (schematically) as follows
\begin{align}
    (cc)_{\bar\three}^1(\c\c)_\three^1\to (c\c)_\eight^S(c\c)_\eight^1\to(c\c)_\eight^S(q\q)_\eight^1\to (c\q)_\one^{S_1}(q\c)_\one^{S_2},
\end{align}
where, as before, the subscripts (superscripts) are colour (spin), and we have suppressed the total $J$ quantum number. In the first step we recouple from the $(cc)(\c\c)$ basis to the $(c\c)(c\c)$ basis, as in the left panel of Fig.\,\ref{decaysfig}, projecting out the colour octet components in which the first pair can have either spin $S=0$ or 1, but insisting that the second pair necessarily has  spin 1 (in order that it can annihilate to a gluon). The second pair then annihilates, via a gluon, to a light $q\q$ pair which is also spin-1 colour-octet. In the final stage we recouple again, projecting out colour singlet $D\*\D\*$ pairs with spins $S_1$ and $S_2$.

The factorisation of the amplitude for the $\f_0'$ component is similar
\begin{align}
(cc)_{\six}^0(\c\c)_{\bar\six}^0\to (c\c)_\eight^S(c\c)_\eight^1\to(c\c)_\eight^S(q\q)_\eight^1\to (c\q)_\one^{S_1}(q\c)_\one^{S_2},
\end{align}
although here the possibilities are fewer, as with total $J=0$ we necessarily have $S=1$ and $S_1=S_2$.

The intermediate step in the above sequences, namely $(c \bar c)_{\eight}^1 \to g \to (q \bar q)_{\eight}^1$, is the same for all transitions, and is independent of the spin $S$ of the spectator $c\c$ pair. When comparing decay rates, its contribution to the amplitude cancels, and so we do not include this factor in our expressions for the amplitude. The remaining colour and spin dependence of the transitions is therefore captured by the recouplings in the first and third step.

With reference to equation~\eqref{colour}, the first colour recoupling contributes a factor $-\sqrt{2/3}$ for $\f_J$ states, and $\sqrt{1/3}$ for $\f_0'$. The colour recoupling in the third step contributes the same factor for all processes, and since this cancels when comparing decay rates, we do not include this factor in our amplitudes.

As for the spin dependence, the numerical factors associated with the first recoupling are those of equation \eqref{fierz}. The recoupling in the third step is a topogically distinct process, with different numerical factors which we summarise here: 
\begin{subequations}
\label{ffierz}
\begin{align}
\|\{(c\c)^1 (q\q)^1 \}^2\> &= \|\{(c \q)^1 (q \c)^1\}^2 \>, \label{ffierz4}\\
\|\{(c\c)^1 (q\q)^1\}^1 \> &= \frac{1}{\sqrt{2}} \|\{(c \q)^0 (q \c)^1\}^1 \> - \frac{1}{\sqrt{2}} \|\{(c \q)^1 (q \c)^0 \}^1\>,\label{ffierz3}\\
\|\{(c\c)^0 (q\q)^1\}^1 \> &= \frac{1}{2} \|\{(c \q)^0 (q \c)^1\}^1 \> + \frac{1}{{2}} \|\{(c \q)^1 (q \c)^0 \}^1\>+\frac{1}{\sqrt 2}\|\{(c \q)^1 (q \c)^1\}^1 \>,\label{ffierz5}\\
\|\{(c\c)^1 (q\q)^1\}^0 \>&= -\frac{\sqrt{3}}{2} \|\{(c \q)^0 (q \c)^0\}^0 \> - \frac{1}{2} \|\{(c \q)^1 (q \c)^1\}^0 \>. \label{ffierz1}
\end{align}
\end{subequations}

Taking $0^{++}\to D\bar D$ as an example, we write the transition amplitude as
\be
\langle  D \D | \hat{H}_2 | 0^{++}\rangle = \phi_{\rm spin} \times  \phi_{\rm colour} \times B(p) \, ,
\ee
where $\phi_{\rm spin}$ and $\phi_{\rm colour}$ are spin and colour matrix elements determined as described above,  and $B(p)$ is the spatial part of the transition amplitude, which we will assume is common to all transitions. For this particular case we find
\be
\langle  D \D | \hat{H}_2 | 0^{++}\rangle = -\left(\frac{1}{2\sqrt 2}\cos\theta+\frac{\sqrt 3 }{4}\sin\theta\right)B(p)
\ee
Equivalent expressions for all of the remaining transitions are in the Appendix.

An interesting feature  is that the ratio of $D\D$ and $D^*\D^*$ amplitudes is the same for both scalars, and is independent of mixing angle,
\begin{align}
  \frac{\<D\D\|\hat H_2\|0^{++}\>}{\<D^*\D^*\|\hat H_2\|0^{++}\>}=
  \frac{\<D\D\|\hat H_2\|0^{++'}\>}{\<D^*\D^*\|\hat H_2\|0^{++'}\>}=\sqrt 3\,,
\end{align}
a result which can be readily understood with reference to equation~\eqref{ffierz1}. It implies that, aside from small differences due to phase space factors, the rates of decay into pseudoscalar and vector meson pairs have the ratio $D\D:D^*\D^* = 3:1$. This applies to quark models (regardless of mixing angle), but notably also applies to the diquark model, which is a special case with $\theta=0$. Working in the diquark model, Ref.~\cite{Becchi:2020uvq} claims the opposite pattern, namely $D\D:D^*\D^* = 1:3$. This incorrect result\footnote{We acknowledge correspondence with Luciano Maiani related to the rate mentioned in Refs.~\cite{Becchi:2020uvq,Becchi:2020mjz}.} also appears in related literature~\cite{Becchi:2020mjz,Maiani:2022psl,Mistry:2023lnm}. 

Taking account of phase space factors, we find
\be
\frac{\Gamma(  0^{++}\to D \bar D)}{\Gamma( 0^{++}\to D^* \bar{D}^*) } ~ \approx ~
\frac{\Gamma(  0^{++'}\to  D \bar D)}{\Gamma( 0^{++'}\to D^* \bar{D}^*) } = 3.12\, ,
\ee
for both quark model mixing ($\theta=35.6\degree$) and the diquark model ($\theta=0$).

\begin{table}
 \renewcommand\arraystretch{1.2}
  \setlength{\tabcolsep}{10pt}
  \centering
\centering
\begin{tabular}{l|ll|ll|ll}
\hline\hline
\multirow{2}*{Final State}&\multicolumn{2}{c|}{$\theta=35.6\degree$} &\multicolumn{2}{c|}{$\theta=0\degree$} & \multirow{2}*{$2^{++}$} & \multirow{2}*{$1^{+-}$} \\
\cline{2-5}
& $0^{++}$ & $0^{++'}$ & $0^{++}$ & $0^{++'}$ &  &  \\ [1ex]
\hline 
$D^* \bar{D}^*$ & 0.14 & 0.011  & 0.062 & 0.094 & 1.0 & 0.248 \\
$D \bar D$ & 0.46 & 0.034  & 0.20 & 0.29 & $\sim 0$ & $-$\\
$D \bar{D}^* + \bar{D} D^*$ & $-$ & $-$  & $-$ &  $-$ & $-$ &   0.252 \\ [1ex]
\hline
\end{tabular}
\caption{The ratio $\Gamma(X\to AB)/\Gamma(2^{++} \to D^* \bar{D}^* )$ for different initial states $X$ and various open-charm final states $AB$, computed as in equation~\eqref{eq:openeg}.}
\label{table:decays2}
\end{table}

For a wider comparison of decays rates for different transitions, we now normalise all decays against the $2^{++}\to D^*\D^*$ mode. As an example we find, from the results in the Appendix,
\begin{align}
    \frac{\Gamma(0^{++}\to D\D)}{\Gamma(2^{++}\to D^*\D^*)}=
    \frac{\omega(0^{++}\to D\D)}{\omega(2^{++}\to D^*\D^*)} \,
    \frac{3}{32}\left(\sqrt 2\cos\theta+\sqrt 3\sin\theta\right)^2,\label{eq:openeg}
\end{align}
where $\omega$ is the relevant phase space factor for the decay. As before, we compute the phase space factors for all decays on the basis of the masses in Table\,\ref{table:masses}.

\begin{figure}
   \centering
    \includegraphics[width=0.8\textwidth]{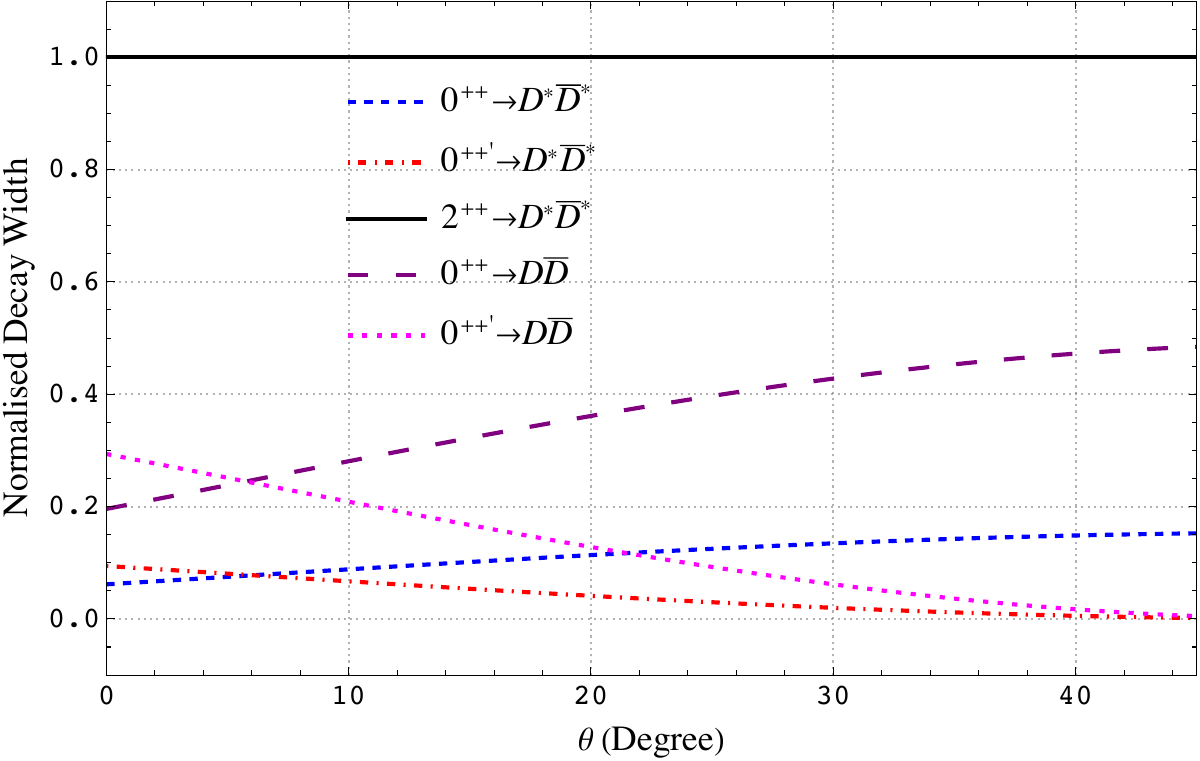}
    \caption{The ratio $\Gamma(X\to AB)/\Gamma(2^{++} \to D^* \bar{D}^* )$ for different initial states $X$ and various open-charm final states $AB$, as a function of the scalar mixing angle $\theta$.}
    \label{mixing2}
\end{figure}

The results obtained in this way 
are shown in Table\,\ref{table:decays2} (for specific values of the mixing angle $\theta$) and Fig.\,\ref{mixing2} (as a function of $\theta$). A notable feature of these results is the dominance of the $2^{++}\to D^*\D^*$ decay in comparison to most other transitions. We therefore suggest the experimental search for $X(6600)$,  which is the tensor state in our scenario, in the $D^*\D^*$ final state. If observed, this channel serves as a benchmark against which other channels can be compared, and confronted with the predictions in Table\,\ref{table:decays2} and Fig.\,\ref{mixing2}. A simple check on our model is that, unlike in $D^*\D^*$, we do not expect a prominent $X(6600)$ signal in $D\D$.

If $X(6600)$ is visible in $D^*\D^*$ then, on the basis of the results in Table\,\ref{table:decays2}, there are good experimental prospects for the discovery of the $1^{+-}$ state in $D^*\D^*$ or $D\D^*/D^*\D$. This is particularly interesting because, as mentioned previously, the mass of the $1^{+-}$ state discriminates strongly between quark and diquark models.

Open-charm decays of the (light) scalar $0^{++}$, which in our scenario is $X(6400)$, are predicted to be somewhat smaller, with a stronger suppression for the diquark model ($\theta=0$) compared to the quark model ($\theta=35.6\degree$). 

As for the heavier scalar $0^{++'}$, the prospects in open charm are not encouraging. In the quark model its decays are strongly suppressed, and in the diquark model this state is absent by construction.

\section{Conclusions}
\label{sec:concl}

There is a growing body of experimental evidence, from LHCb, CMS and ATLAS, for exotic $cc\c\c$ states in the $\jp\jp$ spectrum. We have proposed that two of these states, namely  $X(6600)$ and  $X(6400)$,  belong to an S-wave multiplet of $cc\c\c$ tetraquarks. We have given predictions for their decays in other channels, and additionally have predicted the masses and decays of partner states with other quantum numbers. Many of our predictions can be used to discriminate between competing models, distinguished according to whether quarks or diquarks are the most relevant degrees of freedom.

A simple comparison to the experimental $\Xi_{cc}$ mass, and more detailed model calculations, indicate that the masses of $X(6400)$ and $X(6600)$ are comparable to expectations for the members of an S-wave $cc\c\c$ multiplet. We advocate in particular that $X(6400)$ and $X(6600)$ have scalar ($0^{++}$) and tensor ($2^{++}$) quantum numbers, respectively, because their splitting is then consistent with the predictions of the quark model whose parameters are fixed to the spectrum of ordinary hadrons (see Fig.\,\ref{massplot}). The assignment is also qualitatively consistent with the experimental prominence of the $X(6600)$ peak in $\jp\jp$, relative to $X(6400)$.

By fixing the $X(6400)$ and $X(6600)$ masses to experiment, we can then predict the masses of additional partner states, as shown in Fig.\,\ref{fig:spectrum}. These predictions have either no dependence on model parameters (in the diquark model), or only weak dependence (in the quark model).  A partner state with  axial quantum numbers ($1^{+-}$) is expected in both quark and diquark models, but with a characteristically different mass; as such the discovery of this state can clearly discriminate between models. Another interesting diagnostic would be the discovery (or otherwise) of the heavier scalar ($0^{++'}$), which is expected with a mass around $6600$~MeV in the quark model, but is not expected in the diquark model.
 
We also made predictions for relations among decay branching fractions of $cc\c\c$ tetraquarks to $\jp\jp$, $\jp\eta_c$ and $\eta_c\eta_c$ channels, and among different $D\*\D\*$ channels. 

In the $\jp\jp$ spectrum, in addition to the scalar and tensor states $X(6400)$ and $X(6600)$, in the quark model there is an extra, heavier scalar state, which couples more strong to $\jp\jp$ than the already prominent $X(6600)$. It discovery in this channel would give strong support for the quark model. Lack of signal, conversely, would be less conclusive; it could be that its production is simply suppressed, or, as in the diquark model, that it does not exist.

A very different pattern is expected in the $\eta_c\eta_c$ spectrum. Here we predict a prominent signal only for the scalar $X(6400)$. The tensor $X(6600)$ is not expected to be prominent, as the $\eta_c\eta_c$ channel is a D-wave decay. The additional, heavier scalar state, which is a feature of the quark model only, is not expected to be visible in $\eta_c\eta_c$, as its decay is strongly suppressed by colour mixing. This is one aspect of an interesting pattern in the closed charm decays of $cc\c\c$ states in the quark model: whereas the lowest scalar ($0^{++}$) couples more strongly to $\eta_c\eta_c$ than $\jp\jp$, for the heavier scalar ($0^{++'}$) the pattern is reversed.

The  $\eta_c\jp$ decay mode will be particularly interesting in future experimental studies, as there are good prospects to observe the $1^{+-}$ state, whose mass is a striking diagnostic of the underlying degrees of freedom (quarks versus diquarks).

Among the annihilation decays, we predict that $X(6600)\to D^*\D^*$ is the most significant channel. If observed, this channel sets the scale of annihilation decays, against which other channels can be compared. In particular there would be good prospects for the discovery of the $1^{+-}$ state, which is important for the reason discussed above, in $D^*\D^*$ or $D\D^*/D^*\D$. For the scalar states, the annihilation decays into open charm pairs are predicted to favour $D\D$ over $D^*\D^*$, with relative rates $D\D:D^*\D^* = 3:1$. This applies to both the light scalar ($0^{++}$) in the quark and diquark models, and the heavier scalar ($0^{++'}$) in the quark model, regardless of mixing angle. 

Our predictions for the mass spectrum and decays of $X(6400)$, $X(6600)$ and their possible partners $cc\c\c$ states can ultimately help to distinguish whether quarks or diquarks are the most relevant degrees of freedom for $cc\c\c$ tetraquarks, and are useful to determine their quantum numbers. Once the structure of $cc\c\c$ tetraquarks is understood, it will be helpful to decipher how QCD arranges all-heavy quarks to form exotic hadrons.

\begin{acknowledgments}

We are grateful to Luciano Maiani for his correspondence and to Ryan Bignell for useful discussions and a careful reading of the manuscript. M.\,N.\,A. is grateful to Muhammad Ahmad and Zheng Hu (from CMS Collaboration) for helpful discussions and explanation of the CMS analysis~\cite{CMS:2023owd}, and to the organisers of the LHCb Implication Workshop (in October 2023 at CERN), where this work was presented~\cite{Anwar2023} and experimental prospects discussed. This work is supported by STFC Consolidated Grant ST/X000648/1, and The Royal Society through Newton International Fellowship.

\end{acknowledgments}

\appendix
\section*{Appendix}

The amplitudes for rearrangement decays, obtained as described in Sec.\,\ref{sec:rearr}, are
\begin{align}
    \<\jp\jp\|\hat H_0\|2^{++}\>&=\sqrt\frac{1}{3}A(p)\label{eq:coeffi}\\
    \<\jp\eta_c\|\hat H_0\|1^{+-}\> =\<\eta_c\jp\|\hat H_0\|1^{+-}\> &=\sqrt\frac{1}{6}A(p)\\  
     \<\jp\jp\|\hat H_0\|0^{++}\>&=\left(-\frac{\cos\theta}{2\sqrt 3}+\frac{\sin\theta}{\sqrt 2}\right)A(p)\\
    \<\eta_c\eta_c\|\hat H_0\|0^{++}\>&=\left(\frac{\cos\theta}{2}+\frac{\sin\theta}{\sqrt 6}\right)A(p)\\
     \<\jp\jp\|\hat H_0\|{0^{++}}'\>&=\left(\frac{\sin\theta}{2\sqrt 3}+\frac{\cos\theta}{\sqrt 2}\right)A(p)\\
    \<\eta_c\eta_c\|\hat H_0\|{0^{++}}'\>&=\left(-\frac{\sin\theta}{2}+\frac{\cos\theta}{\sqrt 6}\right)A(p)\label{eq:coefff} \, .
\end{align}
The corresponding amplitudes for annihilation decays (see Sec.\,\ref{sec:anni}) are
\begin{align}
    \<D^*\bar{D}^*|\hat H_2\|2^{++}\>&=-\sqrt\frac{2}{3}B(p)\\
 \< D \bar{D}^*\|\hat H_2\|1^{+-}\> = \<  D^*\D\|\hat H_2\|1^{+-}\> &= -\frac{1}{2\sqrt{3}}~B(p)\\
 \< D^* \bar{D}^*\|\hat H_2\|1^{+-}\> &= -\sqrt\frac{1}{6}~B(p)\\
     \<D^*\bar{D}^*\|\hat H_2\|0^{++}\>&=-\left(\frac{\cos\theta}{2\sqrt 6}+\frac{\sin\theta}{4}\right)B(p)\\
    \<D \D|\hat H_2\|0^{++}\>&=-\left(\frac{\cos\theta}{2\sqrt{2}}+\frac{\sqrt{3}\sin\theta}{4}\right)B(p)\\
     \<D^*\bar{D}^*|\hat H_2\|{0^{++}}'\>&=\left(\frac{\sin\theta}{2\sqrt 6}-\frac{\cos\theta}{ 4}\right)B(p)\\
    \<D \D \|\hat H_2\|{0^{++}}'\>&=\left(\frac{\sin\theta}{2\sqrt{2}}-\frac{\sqrt{3}\cos\theta}{4}\right)B(p) \, .
\end{align}


%

\end{document}